\documentclass[times, twocolumn]{aastex63}
\usepackage{CJK}
\usepackage{graphicx}
\usepackage{enumerate}
\usepackage{amssymb, amsmath}
\usepackage{natbib}
\usepackage{color}
\usepackage{ulem}
\usepackage{dblfnote}
\usepackage{appendix}
\usepackage{hyperref}
\usepackage[stable]{footmisc}
\usepackage{comment}
\usepackage{footnote}
\newcommand{\uat}[2]{\href{http://vocabs.ands.org.au/repository/api/lda/aas/the-unified-astronomy-thesaurus/current/resource.html?uri=http://astrothesaurus.org/uat/#1}{#2 (#1)}}
\usepackage[nodayofweek]{datetime}
\newdateformat{monthyearday}{%
  \THEYEAR\ \monthname[\THEMONTH] \twodigit{\THEDAY}}
  
\submitjournal{AAS Journals}
\shorttitle{Monitoring RY~Tau jet}
\shortauthors{Uyama et al.}

\begin{document}

\title{Monitoring inner regions in the RY~Tau jet}

%% Note that the corresponding author command and emails has to come
%% before everything else. Also place all the emails in the \email
%% command instead of using multiple \email calls.
\correspondingauthor{Taichi Uyama}
\email{tuyama@ipac.caltech.edu}

\author[0000-0002-6879-3030]{Taichi Uyama}
    \affiliation{Infrared Processing and Analysis Center, California Institute of Technology, 1200 E. California Blvd., Pasadena, CA 91125, USA}
    \affiliation{NASA Exoplanet Science Institute, Pasadena, CA 91125, USA}
    \affiliation{National Astronomical Observatory of Japan, 2-21-1 Osawa, Mitaka, Tokyo 181-8588, Japan}

\author[0000-0001-9248-7546]{Michihiro Takami}
    \affiliation{Institute of Astronomy and Astrophysics, Academia Sinica, 11F of Astronomy-Mathematics Building, No.1, Sec. 4, Roosevelt Road, Taipei 10617, Taiwan}

% direct contributions to VAMPIRES/ZIMPOL data

\author[0000-0001-7255-3251]{Gabriele Cugno}
    \affiliation{ETH Zurich, Institute for Particle Physics and Astrophysics, Wolfgang-Pauli-Strasse 27, CH-8093 Zurich, Switzerland}

\author[0000-0003-4514-7906]{Vincent Deo}
    \affiliation{Subaru Telescope, National Astronomical Observatory of Japan, National Institutes of Natural Sciences, 650 North A`oh$\bar{o}$k$\bar{u}$ Place, Hilo, HI 96720, USA}

\author[0000-0002-1097-9908]{Olivier Guyon}
    \affiliation{Subaru Telescope, National Astronomical Observatory of Japan, National Institutes of Natural Sciences, 650 North A`oh$\bar{o}$k$\bar{u}$ Place, Hilo, HI 96720, USA}
    \affiliation{Steward Observatory, University of Arizona, Tucson, AZ 85721, USA}
    \affiliation{Astrobiology Center of NINS, 2-21-1 Osawa, Mitaka, Tokyo 181-8588, Japan}

\author[0000-0002-3053-3575]{Jun Hashimoto}
    \affiliation{Astrobiology Center of NINS, 2-21-1 Osawa, Mitaka, Tokyo 181-8588, Japan}
    \affiliation{Subaru Telescope, National Astronomical Observatory of Japan, Mitaka, Tokyo 181-8588, Japan}
    \affiliation{Department of Astronomy, School of Science, Graduate University for Advanced Studies(SOKENDAI), Mitaka, Tokyo 181-8588, Japan}

\author[0000-0002-3047-1845]{Julien Lozi}
    \affiliation{Subaru Telescope, National Astronomical Observatory of Japan, National Institutes of Natural Sciences, 650 North A`oh$\bar{o}$k$\bar{u}$ Place, Hilo, HI 96720, USA}
    
\author[0000-0002-8352-7515]{Barnaby Norris}
    \affiliation{Sydney Institute for Astronomy, School of Physics, Physics Road, University of Sydney, NSW 2006, Australia}
    \affiliation{AAO-USyd, School of Physics, University of Sydney 2006}

\author[0000-0002-6510-0681]{Motohide Tamura}
    \affiliation{Department of Astronomy, The University of Tokyo, 7-3-1, Hongo, Bunkyo-ku, Tokyo 113-0033, Japan}
    \affiliation{Astrobiology Center of NINS, 2-21-1 Osawa, Mitaka, Tokyo 181-8588, Japan}
    \affiliation{National Astronomical Observatory of Japan, 2-21-1 Osawa, Mitaka, Tokyo 181-8588, Japan}
    
\author[0000-0003-4018-2569]{Sebastien Vievard}
    \affiliation{Subaru Telescope, National Astronomical Observatory of Japan, National Institutes of Natural Sciences, 650 North A`oh$\bar{o}$k$\bar{u}$ Place, Hilo, HI 96720, USA}

% direct contributions to Gemini/Keck data

% from Takami's team (1)

\author[0000-0003-4243-2840]{Hans Moritz G\"unther}
\affil{MIT, Kavli Institute for Astrophysics and Space Research, 77 Massachusetts Avenue, Cambridge, MA 02139, USA}

\author[0000-0002-5094-2245]{P. Christian Schneider}
\affil{Hamburger Sternwarte, Universit\"at Hamburg, Gojenbergsweg 112, D-21029, Hamburg, Germany}

% post-SEEDS LOI and other collaborators
\author[0000-0002-5082-8880]{Eiji Akiyama}
    \affiliation{Department of Engineering, Niigata Institute of Technology, 1719 Fujihashi, Kashiwazaki, Niigata 945-1195, Japan}
    
\author[0000-0002-6881-0574]{Tracy L. Beck}
    \affiliation{The Space Telescope Science Institute, 3700 San Martin Dr., Baltimore, MD 21218, USA}

\author[0000-0002-7405-3119]{Thayne Currie}
    \affiliation{NASA-Ames Research Center, Moffett Blvd., Moffett Field, CA 94035, USA}
    \affiliation{Subaru Telescope, National Astronomical Observatory of Japan, National Institutes of Natural Sciences, 650 North A`oh$\bar{o}$k$\bar{u}$ Place, Hilo, HI 96720, USA}
    \affiliation{Eureka Scientific, 2452 Delmer, Suite 100, Oakland, CA 96002, USA}

\author[0000-0003-0786-2140]{Klaus Hodapp}
    \affiliation{Institute for Astronomy, University of Hawaii, 640 N. Aohoku Place, Hilo, HI 96720, USA}

\author[0000-0003-2815-7774]{Jungmi Kwon}
    \affiliation{Department of Astronomy, The University of Tokyo, 7-3-1, Hongo, Bunkyo-ku, Tokyo 113-0033, Japan}

\author[0000-0002-3424-6266]{Satoshi Mayama}
    \affiliation{The Graduate University for Advanced Studies, SOKENDAI, Shonan Village, Hayama, Kanagawa 240-0193, Japan}

\author[0000-0001-9490-3582]{Youichi Ohyama}
    \affiliation{Institute of Astronomy and Astrophysics, Academia Sinica, 11F of Astronomy-Mathematics Building, No.1, Sec. 4, Roosevelt Road, Taipei 10617, Taiwan}

\author[0000-0002-3273-0804]{Tae-Soo Pyo}
    \affiliation{Subaru Telescope, National Astronomical Observatory of Japan, National Institutes of Natural Sciences, 650 North A`oh$\bar{o}$k$\bar{u}$ Place, Hilo, HI 96720, USA}
    \affiliation{Department of Astronomical Science, SOKENDAI, 2-21-1 Osawa, Mitaka, Tokyo 181-8588, Japan}

\author[0000-0001-9209-1808]{John P. Wisniewski}
    \affiliation{Homer L. Dodge Department of Physics and Astronomy, University of Oklahoma, 440 W. Brooks Street, Norman, OK 73019, US}

%%%%%%%%%%%%%%

\begin{abstract}
We present multi-epoch observations of the RY~Tau jet for H$\alpha$ and [\ion{Fe}{2}] 1.644 \micron~emission lines obtained with Subaru/SCExAO+VAMPIRES, Gemini/NIFS, and Keck/OSIRIS in 2019--2021.
These data show a series of four knots within 1$\arcsec$ consistent with the proper motion of  $\sim$0\farcs3~yr$^{-1}$,  analogous to the jets associated with another few active T-Tauri stars. However, the spatial intervals between the knots suggest the time intervals of the ejections of about 1.2, 0.7, and 0.7 years, significantly shorter than those estimated for the other stars.
These H$\alpha$ images contrast with the archival VLT/SPHERE/ZIMPOL observations from 2015, which showed only a single knot-like feature at $\sim0\farcs25$. The difference between the 2015 and 2019--2021 epochs suggests an irregular ejection interval within the six-year range.
Such variations of the jet ejection may be related to a short-term ($<$1 year) variability of the mass accretion rate.
We compared the peaks of the H$\alpha$ emissions with the ZIMPOL data taken in 2015, showing the brighter profile at the base ($<0\farcs3$) than the 2020--2021 VAMPIRES profiles due to time-variable mass ejection rates or the heating-cooling balance in the jet. 
The observed jet knot structures may be alternatively attributed to stationary shocks, but a higher angular resolution is required to confirm its detailed origin.
\end{abstract}


\section{Introduction} \label{sec: Introduction}

Young stellar objects of various masses and at various evolutionary stages are known to host collimated jets. Those at the base, within 100 au of the star (corresponding to $\sim$$1\arcsec$ for the nearest star forming regions), have drawn particular attentions over many years to understand the jet launching mechanism and its physical link with mass accretion \citep[e.g.][]{Cabrit1990,Hartigan1995,Takami2020}. Unlike more extended parts of the jet, the structures at the base are very compact: in many cases, we have been able to observe them just as chains of knotty structures, as probable signatures of time variable mass ejection, even with the best angular resolutions available \citep[e.g.][for reviews]{Ray2007,Frank2014}. 
Information about more detailed structures in these jets is highly desired to understand the key issues described above.

Furthermore, observations of emission lines at a variety of excitation conditions have been tremendously useful to understand the physical conditions in the jet. 
Such studies have been made for extended energetic jets ($\gg$1000 au from the star) called Herbig-Haro objects \citep[see][for reviews]{Hartigan2000,Reipurth2001}. This contrasts to the observations of jets at the base, which have been made for most cases using only low-excitation forbidden lines like [\ion{O}{1}], [\ion{S}{2}] and [\ion{Fe}{2}] \citep[see][for reviews]{Eisloffel2000,Ray2007,Frank2014}.

State-of-the-art adaptive optics (AO) at optical wavelengths will yield a breakthrough in the studies of jets at the base. 
This technique provides an angular resolution smaller than 0\farcs1 on a 8-10~m telescope, about 2-3 times better than most of the seeing-limited observations to date. 
These instruments enable us to observe H$\alpha$ emission, one of the brightest emission lines associated with the Herbig-Haro objects. This emission line, observed at the base of only a small number of young stars \citep[e.g.][]{Bacciotti2000,Takami2001,Ray2007}, requires an excitation energy significantly larger than the low-excitation needed to trigger the forbidden lines mentioned above, therefore useful for tracing shocks with significantly higher temperatures and/or higher shock velocities etc.

Here we present multi-epoch H$\alpha$ imaging observations of RY~Tau at an extremely high angular resolution (50-60 mas), complemented by near-infrared [\ion{Fe}{2}] observations.
RY~Tau is a T~Tauri Star (TTS) embedded in an envelope and protoplanetary disk in the Taurus star forming region \citep{Kenyon2008}.
The previous measurements of the distance have suggested different values, which makes it hard to accurately characterize the stellar parameters. \cite{Garufi2019} updated the stellar parameters in detail ($T_{\rm eff}$=5750~K, $L_\star=6.3^{+9.1}_{-3.2}\ L_\odot$, $M_\star=1.8\ M_\odot$, and age~$\sim1.8$~Myr) using Gaia DR2-based distance \citep[133~pc;][]{Gaia-DR2}. 
Among pre-main sequence stars, RY Tau hosts one of the most active and best studied jets as highly excited lines such as H$\alpha$ \citep{St-Onge2008,Garufi2019}, \ion{He}{1} \citep{Garufi2019}, and CIV \citep{Skinner2018} have been observed as well as frequently-used forbidden lines such as optical [\ion{O}{1}], [\ion{S}{2}], [\ion{N}{2}] and near-infrared [\ion{Fe}{2}] \citep{Agra-Amboage2009, Garufi2019}.

In this paper, we describe the H$\alpha$ and [\ion{Fe}{2}] data used in this study in Section \ref{sec: Observations and Data Reduction} and their results in Section \ref{sec: Results}. Section \ref{sec: Discussions} discusses substructures in the jet and its movement, as well as the ejection mechanisms.

\section{Observations and Data Reduction} \label{sec: Observations and Data Reduction}

\begin{deluxetable*}{ccccccc}
\label{tab: observations} 
\tablecaption{Observing Logs}
\tablewidth{0pt}
\tablehead{
\colhead{Instrument, Line} & \colhead{Date [UT]} & \colhead{$t_{\rm int}$ [sec]}& \colhead{$N_{\rm exp}$} & \colhead{FWHM [mas]} & \colhead{Mode$^{a}$}  & Field rotation
}
\startdata
        VLT/SPHERE/ZIMPOL, H$\alpha$ & 2015 Nov 8 & 118 &  24 & 50 & SDI & \dots \\
        Geimini/NIFS, [\ion{Fe}{2}] & 2019 Oct 25 & 15 & 108 & 126 & SDI & \dots \\
        Subaru/SCExAO+VAMPIRES$^{b}$, H$\alpha$ & 2020 Jan 31 & 1 & 2497 & 60 & ADI+SDI & 59$^\circ$ \\
        Keck/OSIRIS, [\ion{Fe}{2}] & 2021 Feb 03 & 20 & 46 & 50-100$^c$ & SDI & \dots \\
        Subaru/SCExAO+VAMPIRES$^{b,d}$, H$\alpha$ & 2021 Sep 11 & 0.5 & 6748 & 64 & ADI+double-SDI & 63$^\circ$ \\
\enddata
\tablecomments{$^{a}$ See Section \ref{sec: Observations and Data Reduction} for the detailed explanations of each reduction method.\\
$^{b}$ The VAMPIRES data format is a cube consisting of a 2-dimension image and exposure. The number of exposures ($N_{\rm exp}$) corresponds to the total exposures but in practice we selected 40-50\% percentile in each cube before post-processing. FWHM is measured after the frame selection.\\
$^{c}$ See text for details.\\
$^{d}$ $N_{\rm exp}$ corresponds to the total exposures in State~1 and 2 (see Section \ref{sec: Ha Emission Line}).\\
}
\end{deluxetable*}

\subsection{H$\alpha$ Emission Line} \label{sec: Ha Emission Line}

We observed RY~Tau with Subaru/SCExAO+VAMPIRES on 2020 January 31 UT (engineering run) and on 2021 September 11 (PI: Taichi Uyama, as a backup target).
The detailed exposure settings are presented in Table \ref{tab: observations}.
VAMPIRES enables simultaneous imaging with the narrow-band H$\alpha$ filter ($\lambda_{\rm c}=656.3$ nm, $\Delta\lambda=1.0$~nm) and the adjacent continuum filter ($\lambda_{\rm c}=647.68$~nm, $\Delta\lambda=2.05$~nm) with two detectors. 
To mitigate the aberration between these detectors we utilize a double-differential imaging technique by switching the filters \citep[see][for the schematic]{Norris2015,Uyama2020}.
We set two states where H$\alpha$ and continuum data are taken at the different detectors (State~1: continuum - camera~1, H$\alpha$ - camera~2. State~2 is the other way around). 
This will help to better subtract continuum components from the H$\alpha$ image and we call this subtraction method as `double-SDI' \citep[see below and][for the detailed descriptions]{Uyama2020}.
However, we note that the first VAMPIRES data set, which was obtained for science verification, did not utilize the double-differential imaging technique.

The VAMPIRES data format is a cube consisting of short exposures ($t_{\rm int}$). 
As for data reduction, we first subtract dark from each exposure and conduct PSF fitting by 2D-Gaussian of continuum PSFs for frame selection. 
We then selected the `best’ AO-corrected exposures in each cube based on the peak intensity of the stellar PSF that reflects efficiency of AO correction. We adopted the 40\% and 50\% percentiles, for the first and second data sets respectively, among the fitted PSF peaks as the selection criteria.
We then combined the selected exposures into an image after aligning the centroid of the PSFs \citep[see Figure~2 in][]{Uyama2020}. 
To remove stellar halo and to detect the H$\alpha$ jet feature we applied angular differential imaging \cite[ADI;][]{Marois2006} and spectral differential imaging \citep[SDI; ][]{Smith1987}. 

The preliminary ADI+SDI result for the first VAMPIRES data set is presented in \cite{Uyama2020}, and we updated the post-processing, particularly SDI reduction, as mentioned below.
1) We used the continuum data as a PSF reference for SDI. To prepare SDI reduction we took into account the central star's accretion, difference of the filter transmissions, and the systematic difference of the VAMPIRES detectors. We calculated a scaling factor by comparing photometry (aperture radius = 10$\times$FWHM) of the central star's PSF at each filter. This scaling factor was multiplied to the continuum data before ADI reduction. 
2) We applied ADI reduction to both of the H$\alpha$ and continuum data utilizing {\tt pyklip} algorithms \citep{pyklip}\footnote{\url{https://pyklip.readthedocs.io/en/latest/index.html}}, which produces the most-likely reference PSF for the target PSF by Karhunen-Lo\`eve Image Projection \citep[KLIP:][]{Soummer2012}.
Here we do not conduct aggressive PSF subtraction so that we can avoid severe attenuation and distortion of the jet features. 
We adopted Karhunen-Lo\`eve mode (KL)=1 and minrot=10~deg for the first VAMPIRES data set and KL=3 and minrot=10~deg for the second data set. A larger number of KL allows aggressive speckle suppression while it causes self-subtraction. 
The minrot parameter assumes a frame-to-frame rotation of astrophysical objects and is critical to building PSF reference libraries.
For example, a small minrot value takes nearby pixels of the object into reference PSFs and can cause severe self-subtraction. Note that this parameter does not correspond to the practical frame-to-frame parallactic angle change.
We experimented injection test by burying fake point sources at other position angles of the jet feature and confirmed that these {\tt pyKLIP} settings caused a typical flux loss of $<10\%$ at $\rho>200$~mas and $\sim10-20\%$ between the inner working angle ($\sim100$~mas) and $\rho\geq200$~mas.
3) After the ADI reduction we conducted SDI reduction to the ADI residuals (ADI+SDI) by subtracting the ADI residual image of the scaled continuum data from that of the H$\alpha$ data.

For the VAMPIRES data set taken in 2021, we obtained two states for double-differential imaging. After performing ADI+SDI reduction in each state we applied double-differential imaging to mitigate the aberrations (ADI+double-SDI). 

Finally, we compared the count rate of the unsaturated PSF within an aperture of radius $10\times$FWHM with stellar flux to obtain the count-to-flux conversion factor. 
We utilized the central star's $R$-band magnitude as a photometric reference because we did not observe photometric standard stars.
To take into account variability we referred to the American Association of Variable Star Observers (AAVSO\footnote{\url{https://www.aavso.org/}}). AAVSO did not record the $R$-band magnitude around the VAMPIRES second observing night but the $V$-band magnitude on the nearest date to the second night (September 4 2021) is same as the first night, so we adopted the stellar flux to be 4.3$\times10^{-9}$~erg\ s$^{-1}$cm$^{-2}$\micron$^{-1}$ at 650~nm on both the first and second nights. 
We then converted the post-processed images into the surface brightness H$\alpha$ map.

We also re-reduced the archival VLT/SPHERE/ZIMPOL H$\alpha$ data, which are originally presented in \cite{Garufi2019}. The data from both the B\_Ha ($\lambda_{\rm c}=655.6$~nm, $\Delta\lambda=5.35$~nm) and Cnt\_Ha ($\lambda_{\rm c}=644.9$~nm, $\Delta\lambda=3.83$~nm) filters were reduced with the ZIMPOL pipeline, which rescales the pixels onto a square grid with pixel scale 3.6 mas/pixel and takes care of bias and flat-field corrections. Subsequently, we used {\tt PynPoint} \citep{Stolker2019} to align images fitting a 2D-Gaussian profile to the stellar PSF and to shift the images using spline interpolation. The continuum images were then stretched in the radial direction by the ratio of the filter central wavelengths to align the speckle patterns. Furthermore, they were normalized to the simultaneous H$\alpha$ frame to correct for different band-passes using the counts in an aperture of radius 10$\times$FWHM. At this point, the SDI step could be performed, and we subtracted the continuum images from the H$\alpha$ frames and median combined the residuals, revealing the jets shown in Figure~\ref{fig: result Ha}.
We note that SDI resulted in slight negative sky background and we corrected for it by subtracting the median value of the northern sky area when reporting the results in the following sections.
Finally we converted the SDI result into the surface brightness map as we did in the VAMPIRES data sets. For the ZIMPOL data we followed the prescriptions from \cite{Cugno2019} (see Section 4.1.4 in the literature for detail) to estimate the flux in both the continuum ($F_\mathrm{Cnt\_Ha}=3.55\pm0.11\,\times 10^{-9}$~erg\ s$^{-1}$cm$^{-2}$\micron$^{-1}$) and broad H$\alpha$ ($F_\mathrm{B\_Ha} = 4.09 \pm0.32\times10^{-9}$~erg\ s$^{-1}$cm$^{-2}$\micron$^{-1}$) filters. Briefly, the count rate was estimated in apertures of radius $1\farcs5$ and then converted to physical fluxes using the zero points of the filters estimated in \cite{Schmid2017}.

\subsection{[\ion{Fe}{2}] 1.64-\micron~Forbidden Line} \label{sec: FeII Forbidden Line}
The AO-fed integral field spectroscopic observations of the [\ion{Fe}{2}] 1.644 $\micron$ line were conducted on 2019 October 25 and 2021 February 3, using NIFS at the Gemini North Telescope \citep{NIFS} and OSIRIS at Keck I Telescope \citep{OSIRIS-primary,OSIRIS-Grating}, respectively, as a part of a long-term monitoring program for a few active pre-main sequence stars \citep[][Takami et al., in preparation]{Takami2020}. 
Use of the $H$ and Hn3 gratings with these instruments yielded a spectral resolution $R$$\sim$5500 ($\Delta v$=55 km s$^{-1}$) and $R$$\sim$3800 ($\Delta v$=80 km s$^{-1}$) at 1.5-1.8 and 1.59-1.67 \micron, respectively. These spectral resolutions are optimal to observe the emission lines with high signal-to-noise without resolving their internal kinematics. 

The Gemini data were obtained for a 3\arcsec$\times$3\arcsec filed of view (FoV) through an image slicer with the spatial sampling of 0\farcs1$\times$0\farcs04 along and across the slices, respectively. The Keck data cubes were obtained for a 2\farcs4$\times$3\farcs2 FoV through a lenslet array with a 0\farcs05 spatial sampling. The exposures were made with object-sky-object sequences. The star was placed near the edge of each FoV to cover the (blueshifted) jet with a large spatial area. 

Data were reduced using the pipelines provided by the observatories, and software we developed using PyRAF \citep{pyraf}, numpy \citep{numpy}, scipy \citep{2020SciPy-NMeth} and astropy \citep{astropy:2013,astropy:2018} on python. For the Gemini data, we used the Gemini IRAF package for sky subtraction, flat-fielding, the first stage of bad pixel removals, 2 to 3 dimensional transformation of the spectral data, and wavelength calibration. We then used our own software for stacking data cubes for each date, telluric correction, flux calibration, extraction of the cube for the target emission line, additional removal of bad pixels, and continuum subtraction. We have also corrected a flux loss with the PSF halo, as the jet structures we are interested in are significantly smaller than the PSF halo ($>$0\farcs5). We have used identical processes for the Keck data but data stacking was made using the observatory pipeline. We then integrated the intensity over the velocity range of the emission line to obtain the final images. 

For the Gemini data we measured an angular resolution of 0\farcs12 using the target star in the stacked cube. The target star was saturated in the Keck data, and we estimated an angular resolution of 0\farcs05-0\farcs1 using the snapshots for target acquisition for the target and a spectroscopic standard. The extremely bright stellar continuum compared with line emission causes artifacts due to imperfect continuum subtraction, making the image unreliable within 0\farcs2 of the star. 

\section{Results} \label{sec: Results}

\begin{figure}
\centering
\includegraphics[width=0.45\textwidth]{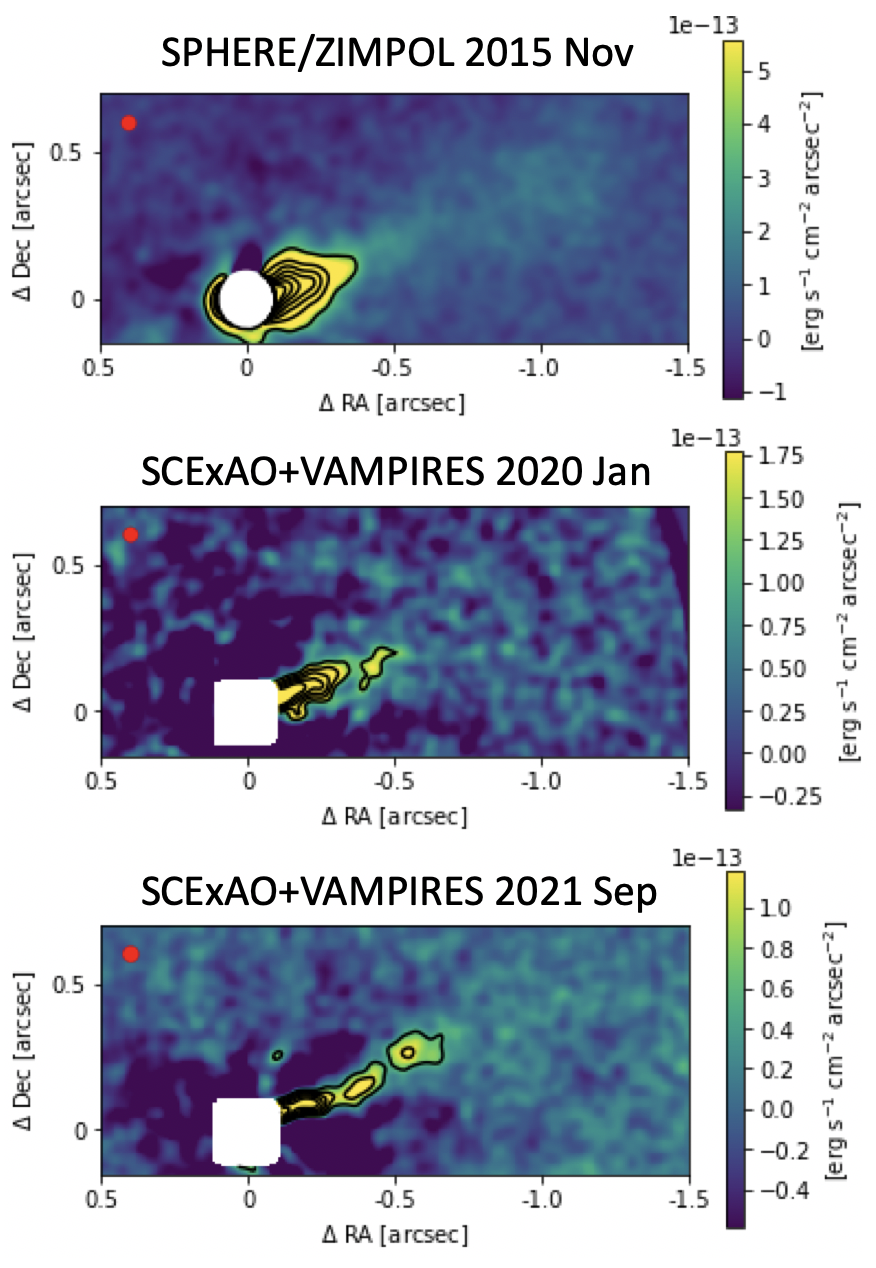}
\caption{Comparison of the VLT/ZIMPOL H$\alpha$ data taken in 2015 November (top) and the Subaru/VAMPIRES H$\alpha$ data taken in 2020 January (middle) and 2021 September (bottom). The images are convolved by 2D-Gaussian (see text) and the color scale and contours are arbitrarily set to clearly show the jet feature. North is up and East is left. The typical FWHM of the observation after convolution is indicated by a red circle at top left of each panel. In the ZIMPOL data we put a $0\farcs1$ mask because the continuum subtraction did not work well.
For the VAMPIRES data sets the central star is masked by the post-processing algorithms. }
\label{fig: result Ha}
\end{figure}

Figure \ref{fig: result Ha} shows the images of the H$\alpha$ jets for three epochs. 
To increase the apparent signal-to-noise, we convolved each of the VAMPIRES and ZIMPOL images 
using a 2D-Gaussian ($\sigma=20$~mas). The FWHM of the PSFs after convolutions increased by $\sim5\%$ from the original unsaturated PSFs.
As a result, the emission was observed for all the epochs with signal-to-noise ratio (SNR) $\gtrsim4$ for the first VAMPIRES data taken in 2020 January and SNR $\gtrsim5$ for the ZIMPOL data and the second VAMPIRES data taken in 2021 September, respectively, at the spine of the jet.
Among three epochs, the ZIMPOL data obtained in 2015 achieved the highest SNR probably thanks to the best AO correction and the brightest H$\alpha$ signal (see Section \ref{sec: Discussions}).
The second VAMPIRES data obtained in 2021 achieved a better SNR than the first VAMPIRES data obtained in 2020 thanks to the double-SDI reduction. 

The H$\alpha$ image in each epoch shows a series of knots at 0\farcs15-0\farcs75 (corresponding to the projected distance of 20-100~au) from the star. The jet structures are different between three epochs, indicative of time variation over six years. Note that the jet features in the VAMPIRES data may be biased in some extent because the jet feature is extended and there could happen a self-subtraction effect though we adopted the non-aggressive ADI parameters as mentioned in Section \ref{sec: Ha Emission Line}. 
The position angle of the jet is roughly measured at $293^\circ$ by eye, which is consistent with previous measurements \citep{Pinilla2018,Garufi2019}, and the latest VAMPIRES data show a curved shape and this feature is possibly connected to the wiggling feature mentioned in \cite{Garufi2019}.

\begin{figure}
\centering
\includegraphics[width=0.45\textwidth]{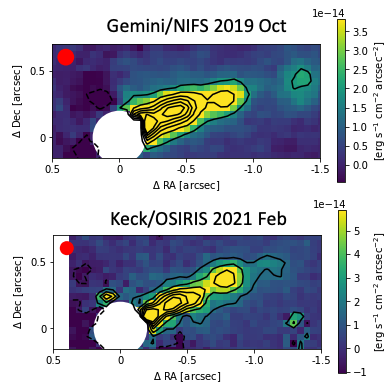}
\caption{Same as Figure \ref{fig: result Ha} but for the Gemini/NIFS [\ion{Fe}{2}] data taken in 2019 October (top) and the Keck/OSIRIS [\ion{Fe}{2}] data taken in 2021 February (bottom). We put a $0\farcs2$ mask in each panel because the continuum subtraction did not work well.}
\label{fig: result Fe II}
\end{figure}

Figure \ref{fig: result Fe II} shows the images of the [\ion{Fe}{2}] jet, which also consists of a series of knots. 
As seen Figures 1 and 2, our H$\alpha$ images appear to show finer structures in the jet probably due to higher angular resolutions. We measure the radial velocity of the jet of $-60$ to $-80$ km s$^{-1}$ in the [\ion{Fe}{2}] emission (with an uncertainty of about $\pm$10 km s$^{-1}$), similar to that measured in the optical [\ion{O}{1}] emission in the past \citep[][]{Agra-Amboage2009}.

\begin{figure}
    \centering
    \includegraphics[width=0.45\textwidth]{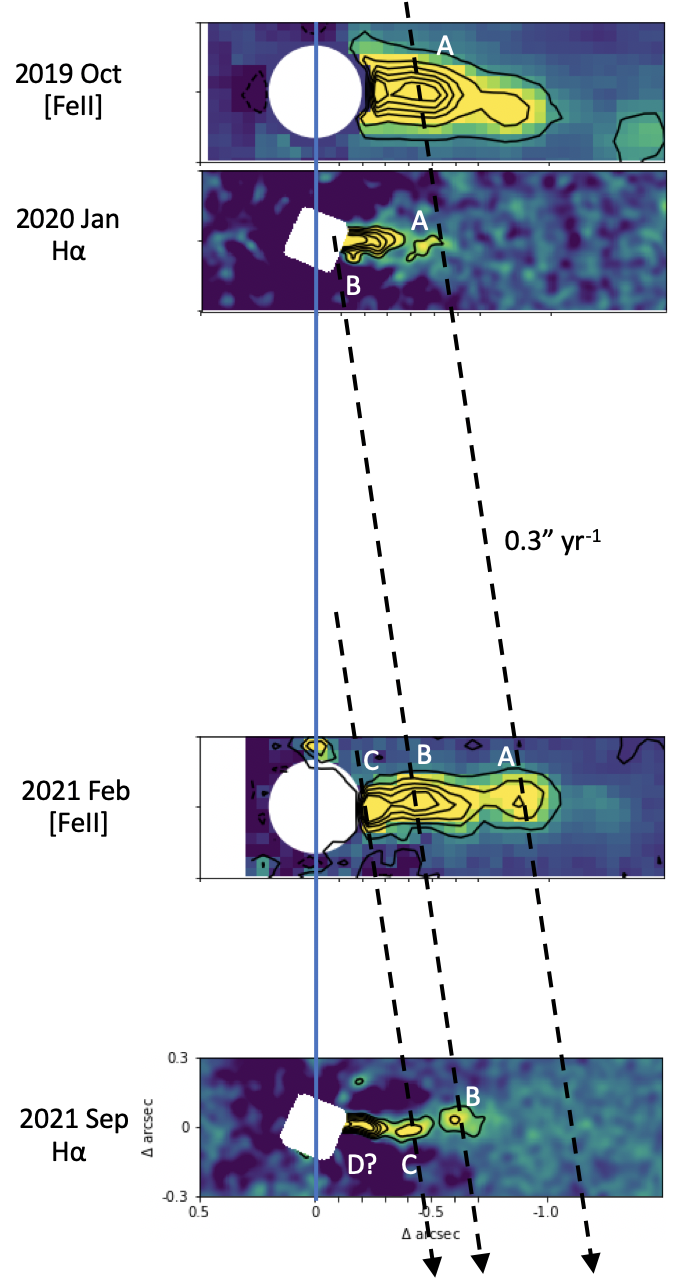}
    \caption{A year-scale monitoring result of the RY~Tau jet. For the clarity we rotated all the field of views by 23$^\circ$ to horizontally align them. The vertical separations between the panels correspond to the time between the epochs. The central star's location is indicated by a dark-blue vertical line. The three dashed lines correspond to roughly-estimated proper motions of the knots A, B, and C ($\sim0\farcs3/{\rm yr}$).}
    \label{fig: jet monitoring}
\end{figure}

To investigate proper motions of the jet knots, we place the H$\alpha$ and [\ion{Fe}{2}] images in 2019--2021 in Figure \ref{fig: jet monitoring}, taking into account the time intervals of the observations.
This figure shows the presence of three or probably four knots and connecting each knot is consistent with a proper motion of $\sim$0\farcs3 yr$^{-1}$.
The proper motion would correspond to a tangential velocity $(V_{\rm T})\sim200\ {\rm km\ s^{-1}}$. \cite{St-Onge2008} monitored the RY~Tau jet at separations $>1\arcsec$ between 1998 and 2004 and measured the tangential velocity as 165~km~s$^{-1}$, and \cite{Garufi2019} suggested $V_{\rm T}\sim$100~km~s$^{-1}$ at separations $<4\arcsec$ and $\sim300$~km~s$^{-1}$ at 5$\arcsec$. The derived value from our observations is roughly consistent with these previous studies and favors a prediction about the tangential velocity from the blue-shifted radial velocity of the jet \citep[][]{Skinner2018}. 
Combining $V_{\rm T}\sim200\ {\rm km\ s^{-1}}$ with the radial velocity measurements from our [\ion{Fe}{2}] observations, we estimate the jet inclination at $\sim70^\circ$, which is consistent with \cite{Agra-Amboage2009}. 
The detailed proper motion and inclination measurements will be presented in the long-term monitoring project paper (Takami et al. in prep).

In Figure \ref{fig: jet monitoring} we mark three knots as A-C and a probable new knot seen at the base of the latest H$\alpha$ image as D (see Section \ref{sec: Discussions} for the detailed discussions).
We note that the H$\alpha$ and forbidden lines reflect different velocity components in terms of quenching and excitation \citep[see e.g.][for details]{Bacciotti2000,Garufi2019} and that H$\alpha$ is more sensitive to the higher velocity than [\ion{Fe}{2}].
Thus, there are possibly slight differences between the peaks of the knots in the H$\alpha$ and [\ion{Fe}{2}] images.

\begin{figure}
\centering
\includegraphics[width=0.45\textwidth]{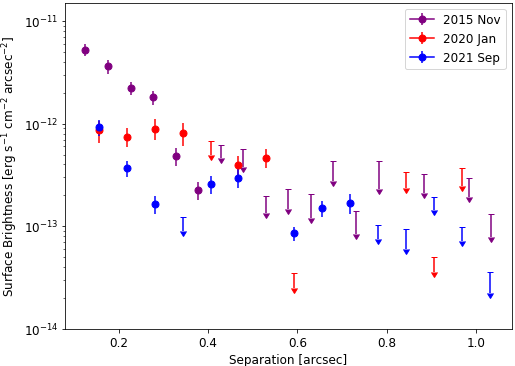}
    \caption{Radial profiles of H$\alpha$ surface brightness from the ZIMPOL (taken in 2015 November) and VAMPIRES data sets (taken in 2020 January and 2021 September). The circle and arrow symbols indicate SNR~$\geq4$ and $<4$ respectively. The SNR is calculated with the {\tt pyklip} cross-correlation functions after 2D-Gaussian convolution as mentioned in Section \ref{sec: Results}. We do not correct flux loss and distortion caused by the post-processing because we adopted the non-aggressive PSF subtraction (see text) and in fact it is difficult to model the H$\alpha$ jet.}
    \label{fig: Ha peak trace}
\end{figure}

Figure \ref{fig: Ha peak trace} shows the traced peak of the H$\alpha$ feature as a surface brightness function of separation. 
We divided the jet feature along separation into several areas by a step of FWHM and adopted the peak pixel as surface brightness at the given separation.
The H$\alpha$ jet observed in 2015 shows brighter surface brightness than that observed between 2020--2021.

\section{Discussion} \label{sec: Discussions}

As shown above, our multi-epoch observations of the jet knots are consistent with the presence of proper motions of $\sim$0\farcs3 yr$^{-1}$. This interpretation is consistent with similar observations for another two TTSs at similar distances ($d$$\sim$140 pc). \citet{Takami2020} presented seven epochs of jet knot imaging in  the [\ion{Fe}{2}] 1.644 $\micron$ emission at $>$0\farcs2 from RW Aur A, and attributed them to moving knots  with proper motions of 0\farcs2-0\farcs3 
yr$^{-1}$. \citet{White2014} combined three epochs of similar observations with those in literature, and identified moving knots at $\gtrsim$0\farcs4 from DG Tau with similar proper motions.

As described in the previous section, the interpretation of moving knots is also consistent with the proper motions of the knots measured downstream ($>$1\arcsec) for the same star. Furthermore, with an estimate of the proper motions of $\sim$0\farcs3, we have derived the jet inclination of $\sim$70\arcdeg, consistent with the previous analysis by another group. Although the present data set does not allow us to accurately estimate the uncertainties of the proper motions, Figure \ref{fig: jet monitoring} suggests that it is significantly smaller than a factor of 2. We believe that this uncertainty does not affect our discussion below.

The X-ray observations of jets from some protostars and young stars indicate the presence of the following two components: (1) an inner stationary component probably related to jet/outflow collimation; and (2) the outer components with proper motions, which are probably related to working surfaces where the shocks travel through the jet \citep[e.g.,][]{Schneider2008,Schneider2011}. Such jets include one associated with DG Tau. The X-ray and near-infrared observations of this jet suggested the location of the stationary component of $\sim$30 au from the star \citep{Gudel2011,White2014}.
Therefore, one may alternatively attribute the observed jet knots shown in Figure \ref{fig: jet monitoring} to stationary shocks, e.g., at 0\farcs2, 0\farcs4 and 0\farcs8 from the star. Our data do not exclude this possibility, however, it might yield the following puzzle: the H$\alpha$ image obtained in February 2021 shows knot B at 0\farcs6 from the star, but none of the images obtained in the other epochs show its counterpart.

For the rest of this section, we {\it tentatively} attribute the jet knots ABC in Figure \ref{fig: jet monitoring} to moving knots as demonstrated. Confirmation of this interpretation requires observations of proper motions with more epochs with smaller time intervals, or those at higher angular resolutions to resolve shock structures.

%Analogous to these jets, bright H$\alpha$ component within 0\farcs3~of the star may be attributed to a stationary shock component, even though Figure \ref{fig: jet monitoring} looks consistent with a proper motion of $\sim$0\farcs3 yr$^{-1}$. 
%Unfortunately, the angular resolutions of our observations are not sufficient for resolving the shock structures and investigating their detailed origin(s). 
%{\bf On the other hand, another system of RW~Aur indicates moving knots at as close as 0\farcs2 to the star \citep{Takami2020}}.
%For the rest of this section, we tentatively attribute the above emission to moving knots as demonstrated in Section \ref{sec: Results}.

To investigate the time variation over a six-year timescale we compared the ZIMPOL data taken in 2015 with the other data sets, particularly with the VAMPIRES H$\alpha$ data. 
In the ZIMPOL data the knot-like feature is detected at $\sim0\farcs25$ while the 2019--2021 data sets show three or four knots within $1\arcsec$.
The difference between the 2015 and 2019-2021 observations suggests that the knot ejection interval is irregular within the six-year range.

In Figure \ref{fig: jet monitoring}, the VAMPIRES H$\alpha$ images show the intervals of the knots A-D of 0\farcs35, 0\farcs2, and 0\farcs2, respectively, indicative of time intervals of the ejections of about 1.2, 0.7 and 0.7 years adopting the proper motion of 0\farcs3 yr$^{-1}$. These intervals are significantly shorter than those measured for a few active T-Tauri jets to date such as DG~Tau \citep[$\sim2.5-5$~yr;][]{Pyo2003,White2014}, RW~Aur \citep[irregular intervals between 3 and 20~yr;][]{Lopez-Martin2003,Takami2020}, and XZ~Tau \citep[$\sim5-10$~yr;][]{Krist2008}.

Such variabilities for jet ejection may be related to activities of mass accretion from the inner disk edge to the star.
Theoretical studies have suggested that the jet ejection mechanism is related to accretion onto the central star \citep[see][for a review]{Najita2000}.
\cite{Petrov1999} and \cite{Grankin2007} monitored RY~Tau's optical spectrum/flux and \cite{Chou2013} investigated mass accretion signatures such as H$\alpha$, which presented short-term variabilities ($\lesssim1$~yr). 
The derived knot interval of 0.7--1.2~years from our results suggests a potential connection with these short-term variability, but we note that the 22d-period variability of H$\alpha$ and Na~D absorption presented by \cite{Petrov2021} is not directly connected to the knot intervals shown in this study.
\cite{Grankin2007} also presented long-term variabilities in the flux of DG~Tau, RW~Aur, and RY~Tau, and \cite{Garufi2019} suggested a possible connection between the RY~Tau jet ejection and an episodic accretion. However RY~Tau does not have unique feature among these three TTS light curves except for circumstellar dust obscuration \citep[e.g.][]{Grankin2007,Petrov2019}.
Detailed investigations of the relationship between the variability and the knot ejection intervals will help to address the RY~Tau jet ejection mechanism in the context of the TTS jet studies.

As shown in Section \ref{sec: Results}, the H$\alpha$ emission at the base ($<$0\farcs3) is brighter in 2015 than 2020--2021 by a factor of 3-6, perhaps due to different mass ejection rates or heating-cooling balance. The latter would be variable due to different shock conditions \citep[see][for a review]{Ray2007} or prompt heating induced by magnetic processes at the base \citep{Skinner2018}. This issue will be discussed in a separate paper with large data sets for jet ejection and signature for mass accretion.

\section{Summary} \label{sec: Summary}

We have presented monitoring results of the RY~Tau jet between 2015 and 2021 using H$\alpha$ and [\ion{Fe}{2}] emission lines with Subaru/SCExAO+VAMPIRES, VLT/SPHERE/ZIMPOL, Gemini/NIFS, and Keck/OSIRIS.
The 2019--2021 data detected a series of three or four knots within 1\arcsec  consistent with the proper motion of $\sim$0\farcs3~yr$^{-1}$, with its uncertainty significantly less than a factor of 2. This would correspond to the tangential velocity of $\sim$200~km~s$^{-1}$, and the radial velocity of $-60$ to $-80$~km~s$^{-1}$. These values and the jet inclination estimated from them are consistent with the previous measurements of the RY~Tau jet. The observed jet intervals between 2019 and 2021 suggest the time intervals of jet knot ejections of about 1.2, 0.7, and 0.7 years and these values are significantly shorter than other T-Tauri jets measured to date.
On the other hand the 2015 data detected only one knot-like feature at $\sim0\farcs25$ and the difference from the 2019--2021 observations suggests the irregular ejection interval within the six-year range.
Such variations of the jet ejection may be related to the short-term variability of mass accretion onto the central star.
We compared the peak of the H$\alpha$ emissions and the 2015 data show the brighter profile at the base ($<0\farcs3$) than the 2020--2021 data perhaps due to different mass ejection rates or heating-cooling balance. This degeneration will be discussed with a large data sets combining jet ejection and mass accretion measurements.

The interpretation of the moving knots within $\sim$100 au of the star is consistent with those in literature for jets associated with another few active T-Tauri stars. Even so, our results do not exclude a possibility that the observed jet knots may be alternatively attributed to stationary shocks related to outflow collimation. However, this scenario would require an explanation about a sudden emergence of one of the knots at 0\farcs6 from the star in the latest epoch of the observations. Throughout, a higher angular resolution is required to confirm the detailed origin of the observed jet knots.

\acknowledgements

The authors would like to thank the anonymous referee for their constructive comments and suggestions to improve the quality of the paper.
T.U. is supported by Grant-in-Aid for Japan Society for the Promotion of Science (JSPS) Fellows and JSPS KAKENHI Grant No. JP21J01220.
M.Takami is supported by the Ministry of Science and Technology (MoST) of Taiwan (grant No. 106-2119-M-001-026-MY3, 109-2112-M-001-019, 110-2112-M-001-044).
G.C. thanks the Swiss National Science Foundation for financial support under grant number 200021\_169131.
M.Tamura is supported by JSPS KAKENHI grant Nos. 18H05442, 15H02063, and 22000005. E.A. is supported by MEXT/JSPS KAKENHI grant No. 17K05399.
H.M.G. was supported by Program number HST-GO-15210.002 provided through a
grant from the STScI under NASA contract NAS5-26555.
P.C.S. gratefully acknowledges support by DLR 50 OR 2102.

This research is based on data collected at the Subaru Telescope, which is operated by the National Astronomical Observatory of Japan, and those via the time exchange program between Subaru and the international Gemini Observatory, a program of NSF’s NOIRLab. 
The part of data presented herein were obtained at the W. M. Keck Observatory, which is operated as a scientific partnership among the California Institute of Technology, the University of California and the National Aeronautics and Space Administration. The Observatory was made possible by the generous financial support of the W. M. Keck Foundation.
The authors wish to recognize and acknowledge the very significant cultural role and reverence that the summit of Maunakea has always had within the indigenous Hawaiian community.  We are most fortunate to have the opportunity to conduct observations from this mountain.
The part of data is based on observations collected at the European Southern Observatory under ESO program 096.C-0454(A).
This work has made use of data from the European Space Agency (ESA) mission
{\it Gaia} (\url{https://www.cosmos.esa.int/gaia}), processed by the {\it Gaia}
Data Processing and Analysis Consortium (DPAC,
\url{https://www.cosmos.esa.int/web/gaia/dpac/consortium}). Funding for the DPAC
has been provided by national institutions, in particular the institutions
participating in the {\it Gaia} Multilateral Agreement.
We acknowledge with thanks the variable star observations from the AAVSO International Database contributed by observers worldwide and used in this research.
This research has made use of NASA's Astrophysics Data System Bibliographic Services.
This research has made use of the SIMBAD database, operated at CDS, Strasbourg, France.

\newpage
\bibliography{library}                                    
%\end{CJK*}
\end{document}